\documentclass[prl,twocolumn,superscriptaddress,amsmath,amssymb]{revtex4}

\usepackage{graphicx}
\usepackage{bm}
\usepackage{amsmath}
\usepackage{amssymb}
\usepackage{amsthm}
\usepackage{multirow}
\usepackage{setspace}

\newcommand {\Fig}[1] {Figure~\ref{#1}}

\newcommand {\tab}[1] {Table~\ref{#1}} 

\newcommand{\beq}{\begin{equation}}
\newcommand{\eeq}{\end{equation}}

\newcommand{\natc}{N@C$_{60}$}

\newcommand{\yatc}{Y@C$_{82}$}
\newcommand{\scatc}{Sc@C$_{82}$}
\newcommand{\laatc}{La@C$_{82}$}

\newcommand{\csixty}{C$_{60}$}

\newcommand{\beqa}{\begin{eqnarray}}
\newcommand{\eeqa}{\end{eqnarray}}

\newcommand{\ttwo}{$T_2$}

\newcommand{\tone}{$T_1$}

\begin{document}

\title{Electron spin coherence in metallofullerenes: Y, Sc and \laatc}

\author{Richard~M.~Brown}
\email{richard.brown@materials.ox.ac.uk} \affiliation{Department of Materials, Oxford University, Oxford OX1 3PH, UK}

\author{Yasuhiro Ito}
\affiliation{Department of Materials, Oxford University, Oxford OX1 3PH, UK}

\author{Jamie Warner}
\affiliation{Department of Materials, Oxford University, Oxford OX1 3PH, UK}

\author{Arzhang Ardavan}
\affiliation{CAESR, Clarendon Laboratory, Department of Physics, Oxford University, Oxford OX1 3PU, UK}

\author{Hisanori Shinohara}
\affiliation{Department of Chemistry and Institute for Advanced Research, Nagoya University, Nagoya 464-8602, Japan}

\author{G.~Andrew.~D.~Briggs}
\affiliation{Department of Materials, Oxford University, Oxford OX1 3PH, UK}

\author{John~J.~L.~Morton}
\affiliation{Department of Materials, Oxford University, Oxford OX1 3PH, UK}
\affiliation{CAESR, Clarendon Laboratory, Department of Physics, Oxford University, Oxford OX1 3PU, UK}

\date{\today}

\begin{abstract}
Endohedral fullerenes encapsulating a spin-active atom or ion within a carbon cage offer a route to self-assembled arrays such as spin chains. In the case of metallofullerenes the  charge transfer between the atom and the fullerene cage has been thought to limit the electron spin phase coherence time (\ttwo) to the order of a few microseconds. We study electron spin relaxation in several species of metallofullerene as a function of temperature and solvent environment, yielding a maximum \ttwo\ in deuterated o-terphenyl greater than 200$\mu$s for Y, Sc and \laatc. The mechanisms governing relaxation (\tone, \ttwo) arise from metal-cage vibrational modes, spin-orbit coupling and the nuclear spin environment. The \ttwo\ times are over 2 orders of magnitude longer than previously reported and consequently make metallofullerenes of interest in areas such as spin-labelling, spintronics and quantum computing.
\end{abstract}

\maketitle

The possibility of encapsulating an atom inside a fullerene cage was discovered by Heath \emph{et al.}~\cite{heath85} in 1985 and has led to widespread research into these novel materials. Encapsulation of a nitrogen atom within a \csixty\ cage (\natc) has been the subject of particular interest due to the remarkably long electron spin coherence times (\ttwo) reported from 80 to 250 $\mu$s~\cite{morton06, morton07}. Metallofullerenes, containing metal ions encased in a similar way, benefit from faster purification and higher production yields than \natc. However, they have not shown particularly long coherence times ($<1.5~\mu$s), attributed to much greater spin density on the fullerene cage~\cite{knorr98, okabe95}. 

The prospect of exploiting the self-assembly of spin-active fullerene molecules into larger structures~\cite{war08} has stimulated interest in spintronics~\cite{lee02} and quantum information processing (QIP)~\cite{fullerene06, har02}. In particular, metallofullerenes can self-assemble within carbon nanotubes in a `peapod' structure, creating a 1-D spin chain~\cite{war08, cantone08, ge08, muj09}. For the potential of such structures to be fully realised, longer electron spin coherence times are required of the constituent metallofullerenes.

Several electron paramagnetic resonance (EPR) studies have been conducted on the metallofullerenes Sc-, Y- and \laatc\, focusing on geometric and electronic properties of the molecules~\cite{ kato93, rubsam95, inakuma00, seifert98, knorr98, okabe95, morley05, ito10}. 
These, along with x-ray diffraction measurements, have shown the metal atom to be off centre in the cage~\cite{nis00, nis98, takata95}, with charge transfer to the cage dependent on the metal ion species~\cite{nag93}. These previous EPR studies have primarily used CW spectroscopy and few \ttwo\ times have have been accurately extracted. Using pulsed EPR, Knorr \emph{et al.}\ report a \ttwo\ of 600 ns for \scatc\ (in trichlorobenzene solvent at 2.5~K) but a measured \ttwo\ of 4.1 $\mu$s in a \yatc\ sample was attributed to a `background' signal~\cite{knorr98}. Okabe \emph{et al.}\  report a temperature and $m_{I}$ dependence of \tone\ and \ttwo\ for \laatc\ (in CS$_{2}$) arising from motional effects of anisotropic interactions and coherence times $<$ 1.5 $\mu$s, in the range 183--283~K~\cite{okabe95}.

In this letter we report pulsed EPR studies of spin relaxation times over a range of temperatures in toluene, deuterated toluene and deuterated o-terphenyl. The latter two are convenient solvents due to their low number of nuclear spins and ability to form a glass at low temperatures, preventing alternative decoherence pathways and clustering (relaxation via dipole-dipole interaction), respectively. We discuss mechanisms governing decoherence (\ttwo) and relaxation (\tone) in for different solvents and metallofullerenes species. We find that under optimised conditions, spin coherence times can exceed previously reported values by over two orders magnitude, rising over 200~$\mu$s.

Experiments were conducted at X-band (9-10~GHz) using a Bruker Elexsys pulsed EPR spectrometer. \tone\ and \ttwo\ were measured by standard inversion recovery ($\pi-\tau-\pi/2-T-\pi-T-echo$) and Hahn echo techniques ($\pi/2-\tau-\pi-\tau-echo$)~\cite{schweiger01} with a $\pi$/2 pulse length of 40~ns. Dilute solutions of Y-, Sc- and \laatc\ ~(10$^{-6}-10^{-7}$ M) were prepared in toluene, deuterated toluene and deuterated o-terphenyl, and degassed using several freeze-pump-thaw cycles. Samples were flash-frozen using liquid N$_{2}$ to produce a glass. $^{45}$Sc and $^{139}$La are both nuclear spin $I=7/2$, while $^{89}$Y has $I=1/2$ (each of these isotopes has 100$\%$ natural abundance). In liquid solution, the isotropic hyperfine structure is clearly visible for all samples but in frozen solutions an anisotropic powder pattern results for Y- and \laatc\ from which individual $m_{I}$ lines cannot be resolved, as previously shown~\cite{knorr98}. In \scatc\ the strong hyperfine coupling can be more clearly resolved~\cite{morley05} and relaxation measurements were conducted on the $m_{I}$=1/2 line, and on the $g_{\perp}$ peak for Y- and \laatc. The broad background as seen by Knorr \emph{et al.}~\cite{knorr98} was not observed. 

The $g$-factors for each metallofullerene were obtained by fitting CW data using Easyspin~\cite{easyspin} giving, ($g_x$, $g_y$, $g_z$)=(2.0022, 2.0023, 1.9974), (1.9969, 2.0036, 1.9999) and (2.0060, 2.0044, 1.9980) for Y-, Sc- and \laatc\ respectively. These are in good agreement with previous studies~\cite{knorr98, morley05}. The hyperfine tensor for \scatc\ is found to be  $A$=[15.4, 7.8, 7.3] MHz in agreement with Morley \emph{et al.}~\cite{morley05} but in contrast to Knorr \emph{et al.}~\cite{knorr98} where hyperfine coupling was not resolved. 

\begin{figure}[t] \centerline
{\includegraphics[width=3.5in]{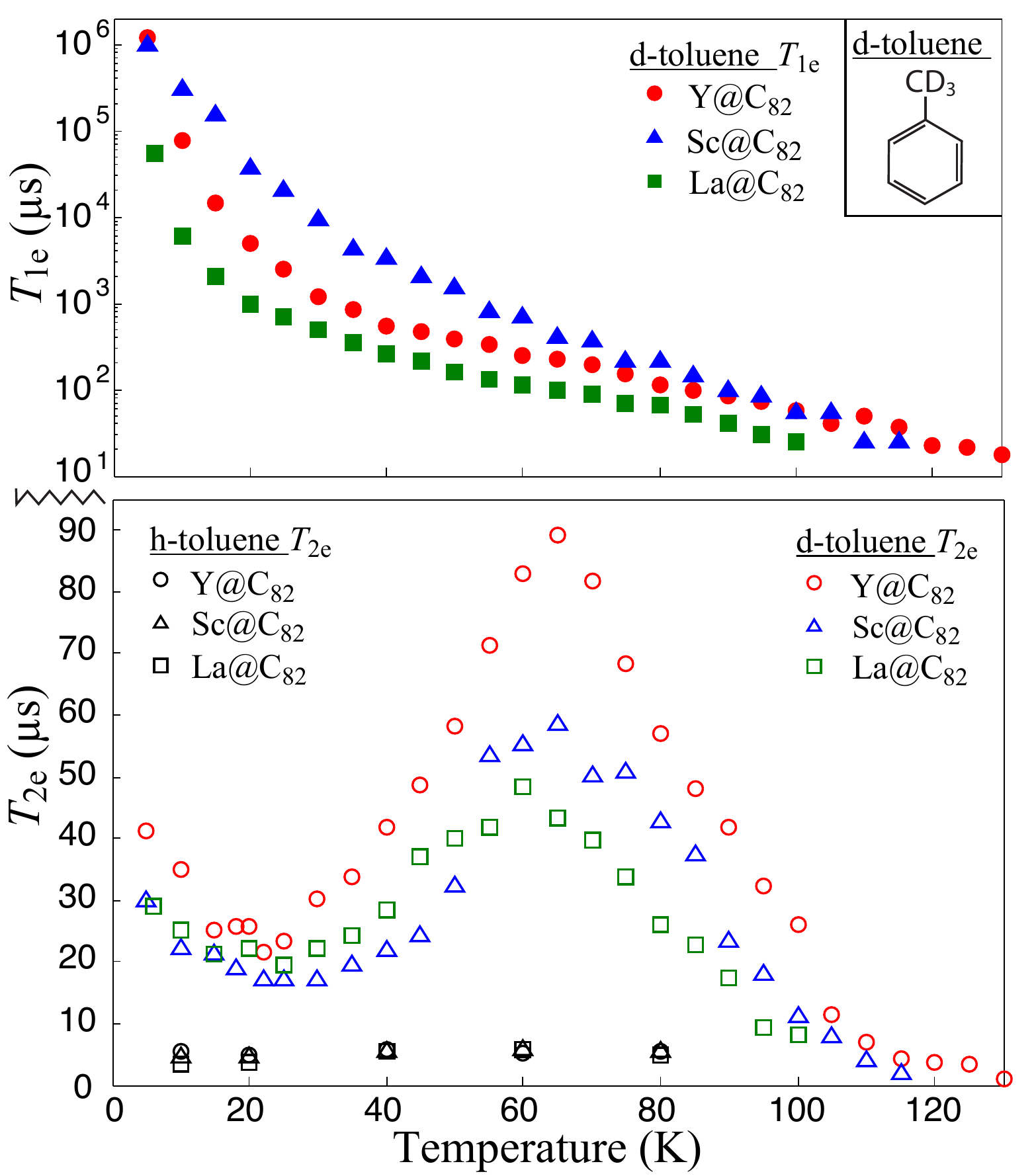}} \caption{{Color online. \tone\ (top, closed shape) and \ttwo\ (bottom, open shape) as a function of temperature for \yatc\ (circle, red), \scatc\ (triangle, blue) and \laatc\ (square, green) in deuterated toluene. \ttwo\ in h-toluene in black shape. \tone\ and \ttwo\ shown on different y scales for clarity. Insert: Structural representation of d-toluene.}} \label{temp}
\end{figure}

\Fig{temp} shows \tone\ and \ttwo\ measured as a function of temperature in h- and d-toluene. \tone\ increases monotonically with decreasing temperature through mechanisms discussed below. \ttwo\ in h-toluene is fairly independent of temperature and limited to $<6\mu$s due to spectral diffusion, a mechanism whereby nuclear spin flips produce a fluctuating local magnetic field that drives electron spin decoherence~\cite{klauder62, zhi69}.
This is identified by a characteristic stretched exponential decay:
\beq
I(t)=I(t_{0}) e^{-(t/T_{2})^{x}}
\eeq
with $x >1$.
This decoherence mechanism is suppressed for metallofullerenes in d-toluene, due to the much weaker nuclear magnetic moment of deuterium. In this deuterated solvent the measured \ttwo\ is longer and a monoexponential decay ($x=1$) is observed for $T>40$~K. \ttwo\ passes through a maximum around 60--70~K of approximately 50, 60 and 90 $\mu$s for \laatc\, \scatc\ and \yatc\ respectively. This is over an order of magnitude longer than previously reported.
The longer coherence times of \yatc\ can be attributed to the weaker hyperfine coupling and smaller nuclear spin of Y compared to Sc and La.

The peak in \ttwo\ at 60--70~K reflects the emergence of an alternative relaxation mechanism independent of the metal ion species, leading to a similar \ttwo\ value for all three metallofullerenes at around 20~K. A prime candidate is the slow rotation of the toluene solvent methyl group which drives deuterium nuclear spin flips and electron spin decoherence via spectral diffusion~\cite{sax97}. The deuterium spin flip rate is maximised at $\omega_{a}^{-1}$=$\tau_{c}$, where $\omega_{a}$ is the deuterium Larmor frequency (2~MHz in the 0.35~T field used here) and $\tau_{c}$ the correlation time~\cite{sax97}. Assuming an Arrhenius temperature dependence for $\tau_{c}$ and an energy barrier equivalent to 270~K~\cite{cav85}, the maximum spin flip rate, corresponding to a minimum in \ttwo, will be at  about 20~K in agreement with experimental results. 

\begin{figure}[t] \centerline
{\includegraphics[width=3.5in]{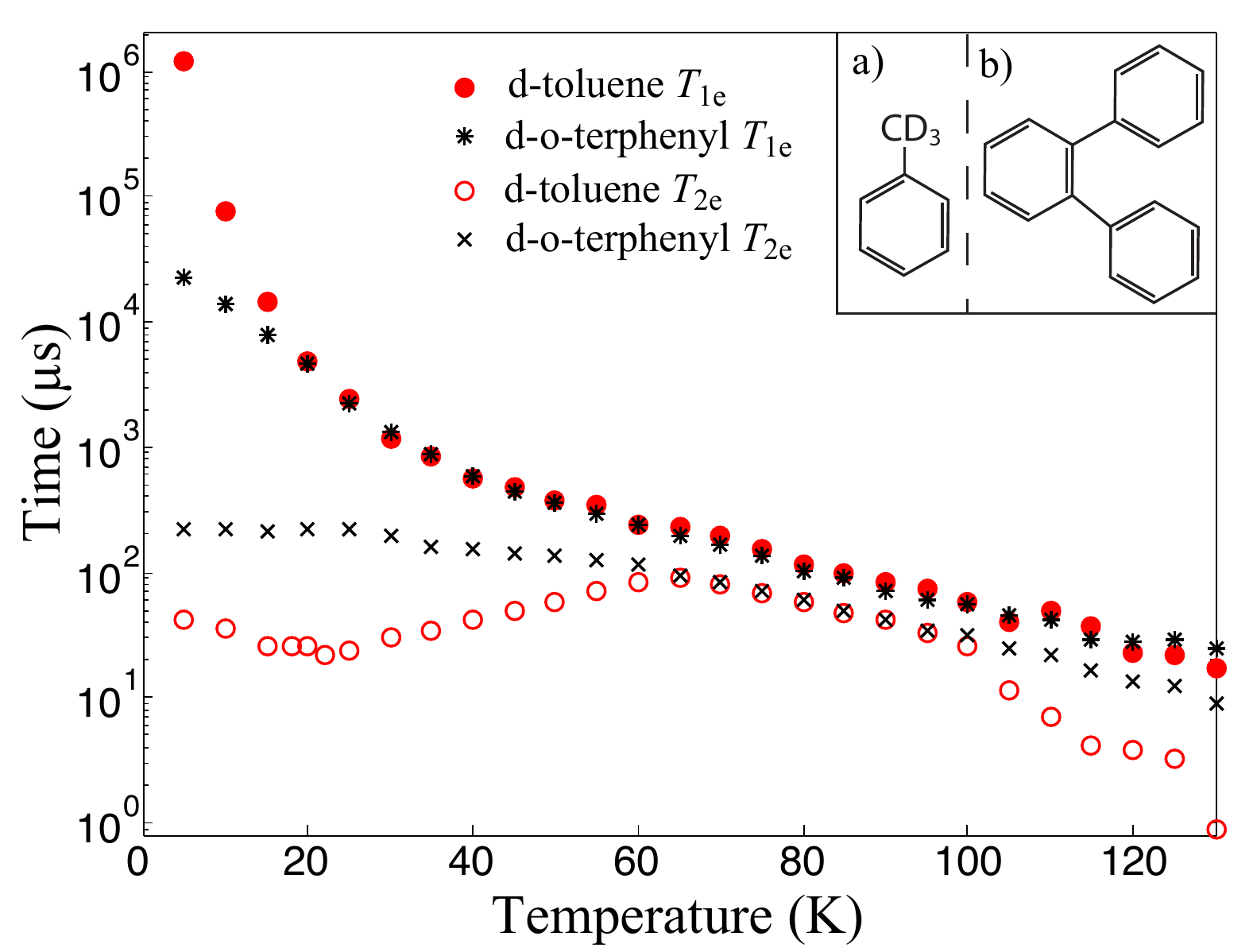}} \caption{{Color online. \yatc\ relaxation and coherence times as a function of temperature in deuterated toluene (circle red, \tone\ closed, \ttwo\ open) and deuterated o-terphenyl (symbol black, \tone\ star, \ttwo\ cross). Insert: Structural representation of a) d-toluene and b) o-terphenyl.}} \label{temp3}
\end{figure}

To overcome this decoherence mechanism and extend \ttwo\ further we chose deuterated o-terphenyl, which contains no methyl groups, as a suitable glass-forming solvent for fullerenes. The \tone\ and \ttwo\ times for \yatc\ in this solvent and d-toluene are compared in \Fig{temp3} (we observed similar results for La and \scatc). \tone\ is largely the same for both solvents, with a deviation at temperatures below 15~K. As predicted, the drop in \ttwo\ observed in d-toluene as the temperature is reduced below 65~K is not observed in deuterated o-terphenyl, due to the absence of methyl groups in this solvent. In the range $T<65$~K, \ttwo~rises slowly with decreasing temperature, and fits well to a stretched exponential in which the stretching factor increases from $x=1$ at 65~K to a limit of $x\approx1.7$ at low temperatures. We attribute this new limit to the emergence of spectral diffusion from the deuterium nuclear spins in the frozen solvent environment. 

In the range 65 to 100~K, \ttwo\ fits to a monoexponential decay and reveals the same coherence times as d-toluene, which appear to be determined by \tone\ relaxation, discussed below. 
Above 100~K, the \ttwo\ values of the two solvents again deviate, which can be associated with the different glass transition temperatures ($T_g$), where $T_g=117$~K for toluene and 243~K for o-terphenyl.

\begin{figure}[t] \centerline
{\includegraphics[width=3.5in]{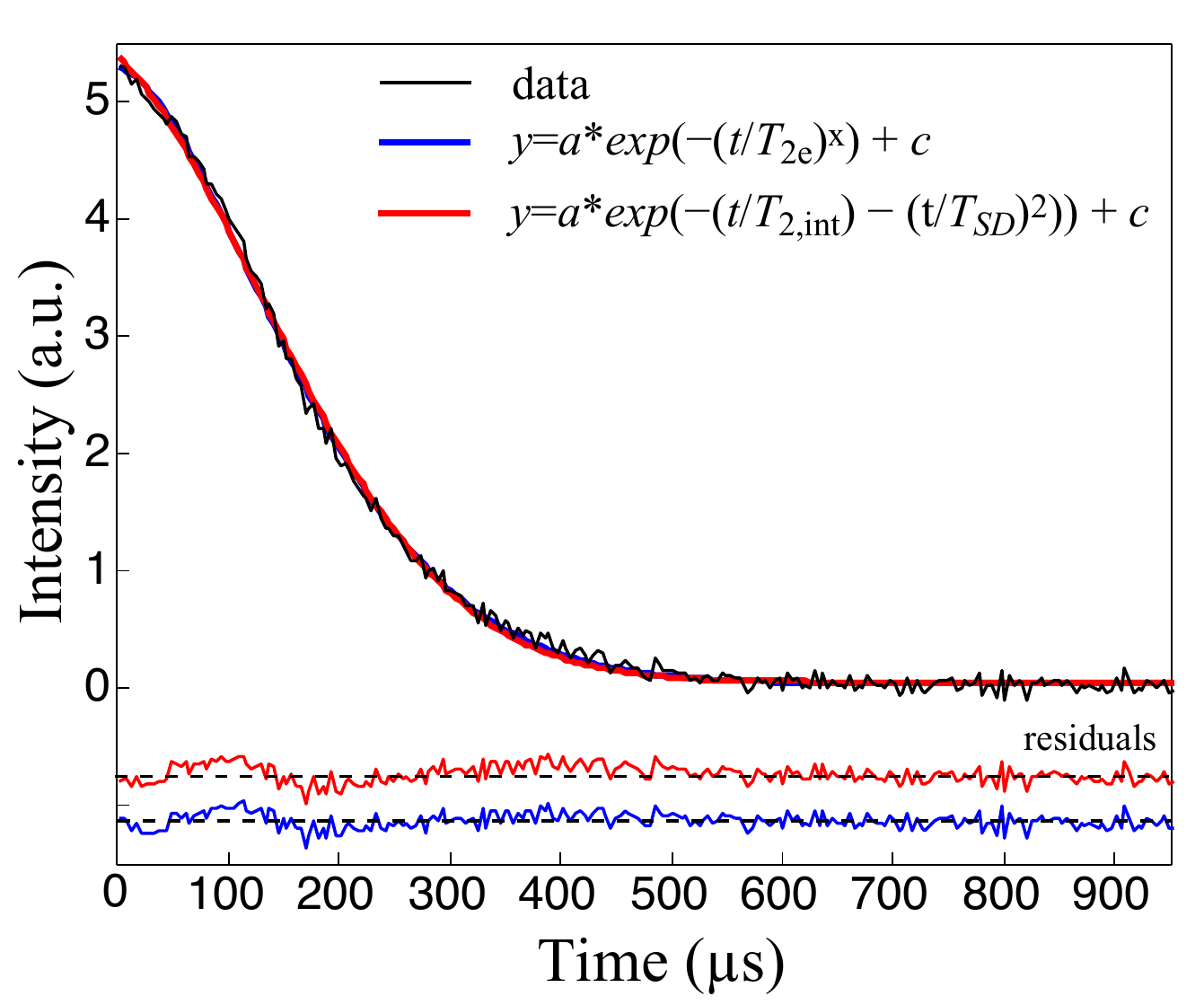}} \caption{{Color online. Hahn echo decay for \laatc\ at 10K, data (black) showing spectral diffusion and fit with a stretched exponential decay (blue) and a curve to extract the intrinsic \ttwo\ in the limit of no environmental nuclear spins (red). Curves are similar and thus overlap, the residual to the fits is shown at the bottom of the figure.}} 
\label{t2}
\end{figure}

In these optimised conditions, we have measured a maximum \ttwo\ of 225$\pm$7, 245$\pm$9 and 204$\pm$2 $\mu$s for \yatc\, \scatc\ and \laatc\ respectively at 5--10~K, with $x=$1.6--1.7$\pm$$<$0.12. A typical electron spin echo decay trace is shown in \Fig{t2}, and fits well to a simple stretched exponential, or a decay of the form:
\beq
I(t)=I(t_{0}) e^{-(t/T_{\rm 2,int})-(t/T_{SD})^{2}}
\eeq
from which an intrinsic $T_{\rm 2,int}$ can be extracted, as the $T_{SD}$ term accounts for spectral diffusion~\cite{zhi69}. This gives $T_{\rm 2,int}= 610\pm 80~\mu$s, offering an estimate of the decoherence time in the absence of spectral diffusion, i.e.\ if all environmental nuclear spins were removed ($T_{SD}= 249\pm 7~\mu$s).

\begin{figure}[t] \centerline
{\includegraphics[width=3.5in]{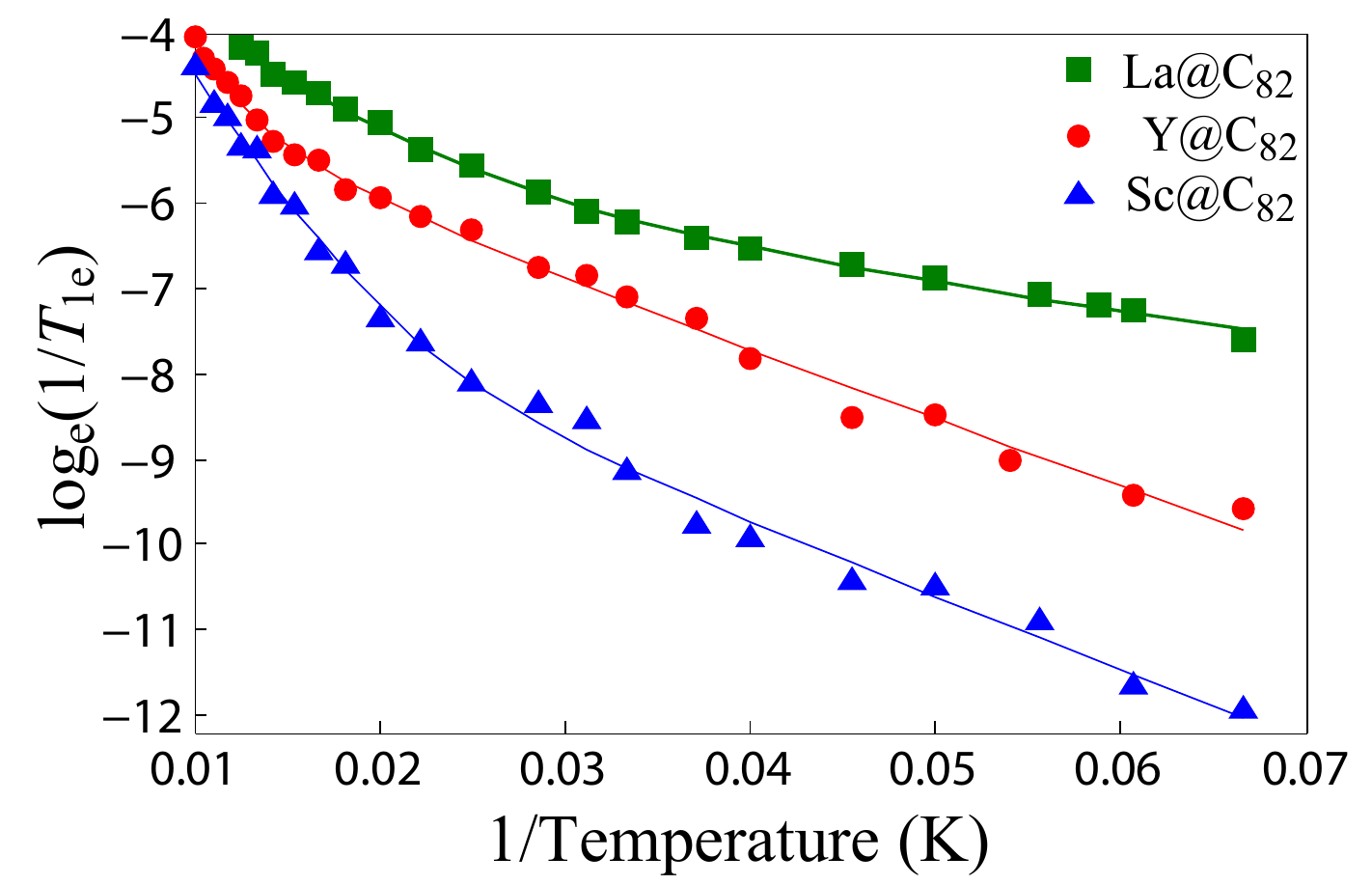}} \caption{{Color online. \tone\ fit to two Arrhenius dependent mechanisms for \laatc\ (square, green), \yatc\ (circle, red) and \scatc\ (triangle, blue). The resonant frequencies extracted from the slope of the lines are shown in~\tab{waven}}} \label{Orbach}
\end{figure}

In contrast to \ttwo, measurements of \tone\ are relatively independent of the solvent environment and show a systematic variation depending on the metal ion species. Furthermore, there is a systematic shift in both the magnitude and temperature dependence of \tone\ which follows the increasing mass of the ions. In the \natc~fullerene, electron spin relaxation has been found to show an Arrhenius-type temperature dependence corresponding to a two-phonon relaxation process via the vibrational motion of the cage~\cite{morton06}. This has the form:
\beq
\label{orb}
T_{1}\propto e^{(\Delta/k_BT)}-1
\eeq
where $\Delta$~is the energy of the molecular vibrational mode. The temperature dependence of \tone\ in Y-, Sc- and \laatc~can be described by two such processes over the range of 15 to 100~K, as shown in~\Fig{Orbach}. The extracted vibrational mode energies (in cm$^{-1}$) for each of the two processes are shown in \tab{waven}.

\begin{table}[h]
\begin{tabular}{cccc}
 Mode  & \multicolumn{3}{c} {Wavenumber (cm$^{-1}$)}\\\
&Y@C$_{82}$ & Sc@C$_{82}$ & La@C$_{82}$ \\\hline
$\nu_{1}$& 54$\pm$4& 59$\pm$8& 19$\pm$5\\
$\nu_{2}$& 403$\pm$214& 229$\pm$43& 125$\pm$33\\
\hline
\end{tabular}
\caption{Extracted molecular vibrational modes from fitting \tone\ temperature dependence in the range 15--100~K, following Eq.~\ref{orb}.}
\label{waven}
\end{table}

The resonant frequencies can be compared to far infrared (FIR) and Raman studies of metallofullerenes that have shown frequencies $<$ 200--300 cm$^{-1}$ to be characteristic of metal-cage vibrations~\cite{lebedkin98, krause02, krause00}. If metal-cage motion is considered in terms of a simple linear oscillator model the vibrational frequency ($\nu$) would be proportional to $\mu^{-1/2}$ (where $\mu$ is the reduced mass). Thus, one would expect the wavenumber to follow La$<$Y$<$Sc as observed for the extracted low frequency mode ($\nu_{1}$) in \tab{waven}. The experimental data will deviate from this model due to, for instance, charge transfer, which is known to be quite different for the metallofullerenes studied ~\cite{nag93, krause02}. However, the general trend holds and therefore is a good indicator that $\nu_{1}$ may be due to metal-cage vibrations. 

Lebedkin \emph{et al.}\ identify a mode for \yatc\ at 54~cm$^{-1}$ which they attribute to a `lateral' metal-cage vibration~\cite{lebedkin98}, in good agreement with the \yatc\ extracted vibrational mode $\nu_{1}$. Similar experimental data on \laatc\ is less conclusive, with a  broader mode centered at $\sim$45--50 cm$^{-1}$ for \laatc~\cite{lebedkin98}, though theoretical studies have predicted `lateral'  La-cage modes at 27 and 30 cm$^{-1}$~\cite{kob03}, slightly higher than our extracted value. This lateral metal-cage mode in \laatc\ has also been identified using x-ray powder diffraction maximum entropy method (MEM) analysis~\cite{nis00}, where a large charge density distribution was attributed to `giant motion'. La would therefore be expected to give a significantly lower metal-cage frequency than from the tight distribution observed for Sc and Y in the C$_{82}$ cage~\cite{nis98, takata95}, consistent with our results. It therefore appears that the lower frequency excited state ($\nu_{1}$) may be due to metal-cage vibrational modes and that this is the dominant cause of \tone\ relaxation in the system over the temperature range 15--60~K.

Above this temperature, the data can be fit by a higher resonant frequency ($\nu_{2}$) (\Fig{Orbach}), though this value is harder to interpret as it may derive from a combination of many higher energy cage vibrational modes. However, $\nu_{2}$ in \laatc\ could correspond to the `longitudinal' metal-cage vibrational mode observed at 163cm$^{-1}$~\cite{lebedkin98}. In addition to the different slopes in the temperature dependence of \tone, the magnitude of \tone\ follows La$>$Y$>$Sc in accordance of the varying extent of charge transfer between the metal and cage~\cite{nag93}. This charge transfer would increase the spin orbit coupling to these molecular vibrational modes, giving the observed order of \tone$^{-1}$ La$>$Y$>$Sc.

In conclusion, we have found long coherence times (\ttwo) over 200 $\mu$s  for all Group III (Sc, Y, La) metallofullerenes limited by spectral diffusion. We have shown \tone\ to depend on metal ion mass and charge state, with evidence that it is driven by metal-cage vibrations. The long coherence times observed, combined with the ability to manipulate spins within tens of nanoseconds put metallofullerenes in a regime where quantum error correction is feasible~\cite{DiV00}. Our identification of the relevant decoherence mechanisms will help inform the choice of structure for larger fullerene arrays. In addition to potential applications in quantum information and spintronics, these long coherence times could make metallofullerenes a candidate EPR spin label.

We thank Gareth and Sandra Eaton for helpful discussions. The research is supported by the EPRSC through the QIP IRC (No. GR/
S82176/01 and GR/S15808/01)), and CAESR (No. EP/D048559/1). R.M.B is supported by EPRSC and project IMPRESS, J.J.L.M and A.A are supported by the Royal Society and G.A.D.B by EPRSC.

\bibliography{bib5}
\end{document}